\documentclass{blois}

\usepackage{amsmath}
\usepackage{slashed}
\usepackage{amssymb}
\usepackage{color}
\usepackage[table]{xcolor}

\def\be{\begin{equation}}
\def\ee{\end{equation}}
\def\bea{\begin{eqnarray}}
\def\eea{\end{eqnarray}}

\newcommand{\footstar}[1]{$^*$ \footnotetext{$^*$#1}}

\def\fig#1{{Fig.~\ref{#1}}}

\def\Table#1{{Table~\ref{#1}}}

\definecolor{LightCyan}{rgb}{0.88,1,1}
\definecolor{piggypink}{rgb}{0.99, 0.87, 0.9}
\definecolor{applegreen}{rgb}{0.55, 0.71, 0.0}
\definecolor{darkpastelgreen}{rgb}{0.01, 0.75, 0.24}
\definecolor{green-yellow}{rgb}{0.68, 1.0, 0.18}

\newcommand{\beq}{\begin{equation}}
\newcommand{\eeq}{\end{equation}}
\newcommand{\Photo}{}

\begin{document}
\vspace*{4cm}
\title{ACCIDENTALLY SAFE EXTENSIONS OF THE STANDARD MODEL}

\author{L.~DI LUZIO$^1$\footstar{Talk given at the 27th Rencontres the Blois on Particle Physics and Cosmology, 
May 31 - June 05, 2015.
%Speaker
}, R.~GR\"OBER$^2$, J.~F.~KAMENIK$^{3a,b,c}$, M.~NARDECCHIA$^4$}

\address{
$^1$Dipartimento di Fisica, Universit\`a di Genova and INFN, Sezione di Genova, \\ via Dodecaneso 33, 16159 Genova, Italy \\
$^2$INFN, Sezione di Roma Tre, via della Vasca Navale 84, 00146 Roma, Italy \\
$^{3a}$Jo\v{z}ef Stefan Institute, Jamova 39, 1000 Ljubljana, Slovenia, \\
$^{3b}$Faculty of Mathematics and Physics, University of Ljubljana, Jadranska 19, 1000 Ljubljana, Slovenia \\
$^{3c}$CERN TH-PH Division, Meyrin, Switzerland \\
$^4$DAMTP, University of Cambridge, Wilberforce Road, Cambridge CB3 0WA, United Kingdom
}

\maketitle\abstracts{
We discuss a class of weak-scale extensions of the Standard Model which is completely invisible to low-energy indirect probes. 
The typical signature of this scenario is the existence of new charged and/or colored states which are stable on the scale of high-energy particle detectors.}

\section{Introduction}

\begin{figure}[ht]
\centering
\includegraphics[angle=0,width=15cm]{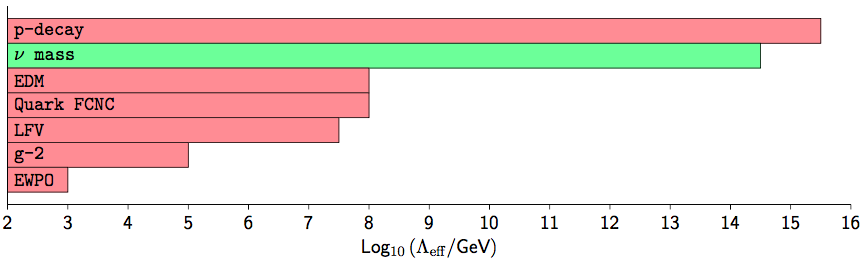}
\caption{\label{fig:IB} Summary of indirect bounds on the scale of new physics (NP). Massive 
neutrinos hint to a generic high-scale dynamics which automatically accommodates all 
indirect searches for NP. A future improvement on proton decay bounds and/or failure 
to observe neutrinoless double beta decay 
with inverse neutrino mass hierarchy would put pressure on this framework.}
\end{figure}

Consider the Standard Model (SM) as the renormalizable part of an effective field theory (EFT)
\begin{equation}
\mathcal L = \mathcal L_{\rm SM}^{(d\leq 4)} + \sum_{d>4} \frac{1}{\Lambda_{\rm eff}^{d-4}} \mathcal L^{(d)} \,,
\end{equation}
where only SM fields appear as dynamical degrees of freedom in $\mathcal L$, and 
$d$ denotes the canonical operator dimension. Assuming $\mathcal O(1)$ coefficients in the 
EFT operator expansion, currently all experimental evidence in particle physics can be accommodated 
by such a generic theory with a very large cut-off scale $\Lambda_{\rm eff} \approx 10^{15}$~GeV (cf.~\fig{fig:IB}).

For NP around the TeV scale to be compatible with the indirect bounds 
in \fig{fig:IB}, exact or approximate symmetries are usually postulated
which in term forbid or sufficiently suppress the most dangerous contributions to 
flavor changing neutral currents, CP violation, as well as B and L changing processes. 
At the heart of these problems is the fact that generically, extending 
the SM particle content will either break some of the SM accidental symmetries, 
and/or introduce new sources of breaking of the approximate SM symmetries, 
which in general will not commute with the existing SM spurions. 
Examples of the first kind include B and L. Flavor, 
CP and custodial symmetry of the Higgs potential fall into the second category. 

One may thus ask the following well defined question: 
{\it Which extensions of the SM particle content with masses close to the 
electroweak scale (i) automatically preserve the accidental and approximate 
symmetry structure of the SM, 
(ii) are cosmologically viable, 
and (iii) form consistent EFTs with a cut-off scale 
as high as $10^{15}$~{\rm GeV}?}

In Ref.~\cite{DiLuzio:2015oha} we explored such possibility by adding to the SM one extra matter (spin $0$ or $1/2$) 
multiplet and requiring that its SM gauge quantum numbers alone forbid all renormalizable interactions which would 
break any of the SM approximate or accidental symmetries. 
In the present contribution we review 
the selection criteria which lead to the answer to the question above. 

\section{Accidental matter multiplets}

The classification of single particle SM extensions which automatically preserve 
its accidental and approximate symmetry structure leads to an infinite set of states. 
In order to make the list finite we apply two extra selection 
criteria coming from cosmology (no too long-lived colored/charged particles) 
and perturbativity (no Landau poles below $\Lambda_{\rm eff} \approx 10^{15}$~{\rm GeV}). The final list of viable accidental matter multiplets which pass all the 
constraints is provided in \Table{summary1} (uncolored case). This Table is complemented by an analogous 
one featuring $14$ ($3$) colored 
scalars (fermions) which can be found in Ref.~\cite{DiLuzio:2015oha}. 

\setlength{\tabcolsep}{4pt}

\begin{table}[htbp]
\centering
\caption{\label{summary1} 
  List of new weak-scale uncolored states $\chi$ which can couple to SM fields 
  at the renormalizable level
  without breaking the flavor group $U(3)^5$, 
  and which are compatible with cosmology and  
  an EFT cut-off scale of $\Lambda_{\rm{eff}} \approx 10^{15}$ GeV. 
  The possible electromagnetic charges of the LP in the multiplet are denoted by $Q_{\rm{LP}}$, while 
  $\mathcal{O}_{\rm decay}$ denotes the lowest dimensional operators responsible for the decay of $\chi$. 
  States with $Y=0$ are understood to be real.
  In the last column, the Landau pole has been 
  estimated at two loops by integrating in the new multiplet at the scale of the $Z$ boson mass, while the symbol in the bracket stands for the gauge coupling, $g_{1,2,3}$, triggering the Landau pole and $m_{\rm Pl} = 1.22 \times 10^{19}$~GeV is the Planck mass. The states marked with $\ddag$ and $\star$ are constrained by electroweak precision tests and BBN, respectively, to  lie possibly beyond the LHC reach.}
\vspace{0.4cm}
\begin{tabular}{@{} |c|c|c|c|c|c| @{}}
\hline
Spin  & $\chi$ &  $Q_{\rm{LP}}$ & $\mathcal{O}_{\rm decay}$ & dim$(\mathcal{O}_{\rm decay})$ & $\Lambda_{\rm{Landau}}^{\rm{2-loop}}$[GeV] \\ 
\hline
%\hline
0 & $(1,1,0)$ & 0  & 
$\chi H H^{\dag}$  & 3 &$\gg m_{\rm Pl}$ ($g_1$)  \\ 
0 & $(1,3,0)^\ddag$ & 0,1 & $ \chi H H^{\dag} $ & 3 & $\gg m_{\rm Pl}$ ($g_1$)  \\ 
0 & $(1,4,1/2)^\ddag$ & -1,0,1,2 & $ \chi H H^{\dag} H^{\dag} $ & 4 & $\gg m_{\rm Pl}$ ($g_1$)  \\ 
0 & $(1,4,3/2)^\ddag$ & 0,1,2,3 & $ \chi H^{\dag} H^{\dag} H^{\dag} $ & 4 & $\gg m_{\rm Pl}$ ($g_1$)  \\ 
\hline
\rowcolor{LightCyan}
0 & $(1,2,3/2)$ & 1,2  & $\chi H^{\dag} \ell \ell ,~ \chi^{\dag} H^{\dag} e^c e^c ,~ D^\mu \chi^\dag \ell^\dag \overline{\sigma}_\mu e^c$  & 5 &$\gg m_{\rm Pl}$ ($g_1$)  \\ 
\rowcolor{LightCyan}
0 & $(1,2,5/2)$ & 2,3 & $ \chi^{\dag} H e^c e^c $ & 5 & $\gg m_{\rm Pl}$ ($g_1$)  \\ 
\rowcolor{LightCyan}
0 & $(1,5,0)$ &  0,1,2  & $ \chi H H H^{\dag} H^{\dag} ,~ \chi W^{\mu \nu} W_{\mu \nu} ,~ \chi^3  H^{\dagger} H $ & 5 & $\gg m_{\rm Pl}$ ($g_{1}$) \\ 
\rowcolor{LightCyan}
0 & $(1,5,1)$ &  -1,0,1,2,3 & $ \chi^{\dag} H H H H^{\dag},~\chi \chi \chi^{\dagger} H^{\dagger} H^{\dagger} $ & 5 & $\gg m_{\rm Pl}$ ($g_1$) \\ 
\rowcolor{LightCyan}
0 & $(1,5,2)$ & 0,1,2,3,4 & 
$ \chi^{\dag} H H H H$ & 5 &  $3.5 \times 10^{18}$ ($g_1$)  \\ 
\rowcolor{LightCyan}
0 & $(1,7,0)^\star$ & 0,1,2,3 & $\chi^3 H^{\dagger} H$
& 5 & $1.4 \times 10^{16}$ ($g_{2}$)  \\ 
\hline
\rowcolor{piggypink}
1/2 & $(1,4,1/2)$ &  -1 & $\chi^c \ell H H,~\chi \ell H^{\dag} H ,~ \chi \sigma^{\mu \nu} \ell W_{\mu \nu}$ & 5 & $8.1 \times 10^{18}$ ($g_2$)  \\ 
\rowcolor{piggypink}
1/2 & $(1,4,3/2)$ &  0 & $\chi \ell H^{\dag} H^{\dag}$  & 5 & $2.7 \times 10^{15}$ ($g_1$) \\ 
\rowcolor{piggypink}
1/2 & $(1,5,0)$ &  0 & $\chi \ell H H H^{\dag} ,~ \chi \sigma^{\mu \nu} \ell H W_{\mu \nu}$  & 6 & $8.3 \times 10^{17}$ ($g_2$) \\ 
  \hline
  \end{tabular}
\end{table}

\subsection{Automatic preservation of $U(3)^5$}

The quantum numbers of the new multiplet $\chi$ are selected by 
requiring that it does not couple to SM 
fermions at the renormalizable level. 
This automatically guarantees that the flavor group $U(3)^5$ 
of quarks and leptons (including B and family L number)
is only broken by their respective Yukawas. 

Let us consider first the case where $\chi$ is a fermion. 
If $\chi$ transforms under a real representation of the SM gauge group, the most general 
Lagrangian is 
\begin{equation}
\mathcal{L}= \mathcal{L}_{\text{SM}} + i \chi^{\dagger} \overline{\sigma}^{\mu} D_{\mu} \chi + \frac{1}{2} M (\chi^T \epsilon \chi + \text{h.c.}) \, ,
\end{equation}
which is invariant under a $Z_2$ transformation $\chi \to -\chi$. 
If $\chi$ transforms instead under a complex or pseudoreal representation of the gauge group (so that a Majorana mass term is forbidden), we introduce another field $\chi^c$ with conjugate quantum numbers. In this way, the new state is vector-like and a mass term can always be added. In both cases 
an extra accidental symmetry guarantees the stability of the new particle at the renormalizable level. The case of extra scalars is more involved since they can always couple to the Higgs field at the renormalizable level 
without breaking $U(3)^5$ 
and their stability depends on the allowed interactions with the Higgs field (cf.~Ref.~\cite{DiLuzio:2015oha}).

\subsection{Cosmology}

Barring few exceptions (for scalar cases only), the lightest particle (LP) in the multiplet is stable at the renormalizable level. Colorless and electrically neutral stable particles can be 
dark matter (DM) candidates, a possibility which has been extensively studied in the literature, also known as Minimal DM~\cite{Cirelli:2005uq}. On the other hand, colored and charged stable particles are severely constrained by cosmological observations as well as by searches for exotic forms of matter on Earth and in the Universe \cite{Burdin:2014xma}. 

The new extra states will eventually decay due to operators present in the 
SM$+\chi$ EFT  
suppressed by the scale $\Lambda_{\rm eff} \approx 10^{15}$~{\rm GeV}. 
Since we are interested in LHC-scale extensions of the SM 
we can set $m_\chi \sim 1$ TeV, and estimate on dimensional grounds the lifetime 
of the particle due to $d=5,6,\ldots$ operators: 
\begin{equation}
\Gamma_5 \sim \frac{m^3_\chi}{\Lambda_{\rm eff}^2} \approx (10^{-3} \ s)^{-1} \, ,
\qquad
\Gamma_6 \sim \frac{m^5_\chi}{\Lambda_{\rm eff}^4} \approx (10^{21} \ s)^{-1} \, , 
\qquad \ldots \, .
\end{equation} 
Particles decaying via $d=5$ operators are potentially subject to Big Bang Nucleosynthesis (BBN) constraints. A refined calculation of the particle lifetimes and the 
relic abundances shows that BBN bounds are indeed relevant for some of the 
new states in \Table{summary1} (cf.~Ref.~\cite{DiLuzio:2015oha} for more details). 
On the other hand, charged and/or colored particles decaying via $d>5$ operators 
are excluded by cosmic rays and exotic heavy nuclei searches \cite{Burdin:2014xma}.

\subsection{Landau poles}

Extra matter drives the gauge couplings towards the non-perturbative regime and 
can eventually lead to the emergence of a Landau pole. The latter might be associated 
with a new dynamics that, if generic, would break the accidental symmetries of the SM. 
We hence require that no Landau poles are generated below the cut-off of the EFT, 
$\Lambda_{\rm eff} \approx 10^{15}$ GeV. This yields in turn an upper bound on the dimensionality of the extra representation. 

\begin{figure}[ht]
\centering
\includegraphics[angle=0,width=8.5cm]{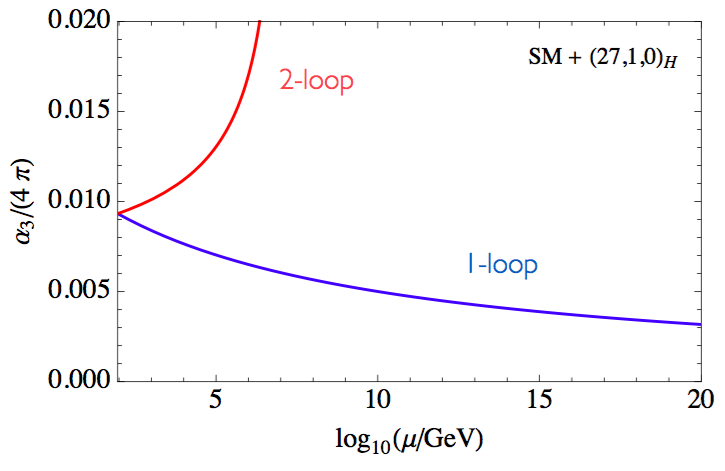}
\caption{\label{fig:2loopLP} 
Running of $\alpha_3$ for the SM augmented with a real scalar multiplet 
with quantum numbers $(27,1,0)$
under $SU(3)_C \otimes SU(2)_L \otimes U(1)_Y$.
}
\end{figure}

The determination of the Landau pole is carried out at the two-loop level. 
Indeed, for non-abelian gauge factors there can be an accidental cancellation in the one-loop beta function 
between matter and gauge contributions, so that two-loop effects may become important and even change qualitatively the asymptotic behavior of the 
theory (cf.~\fig{fig:2loopLP}).

\section{Conclusions}

Many indirect probes of NP such as proton decay, flavor and CP violating observables tell us that NP 
at the TeV scale (if there) must be highly non-generic. In order to suppress 
the most dangerous $d=6$ operators in the SM EFT, extra protection 
mechanisms like MFV~\cite{D'Ambrosio:2002ex}, B and/or L number conservation are usually imposed. In the work in Ref.~\cite{DiLuzio:2015oha}, 
we explored yet another possibility: 
\emph{the gauge quantum numbers of the NP states are such that the accidental and approximate 
symmetry structure of the SM is automatically preserved at the renormalizable level}.  
After applying cosmological constraints and by requiring the consistency 
of the SM$+\chi$ EFT up to $\Lambda_{\rm eff} \approx 10^{15}$ GeV we are left with a finite set of 
single particle extensions of the SM (cf.~\Table{summary1}), 
which were the focus of a detailed phenomenological analysis including 
BBN constraints and collider bounds. 
Interestingly, these scenarios feature in most of the cases a striking phenomenological 
signature with charged and/or colored states which are stable on the scale 
of high-energy particle detectors. 

There are also a couple of side products of the analysis in Ref.~\cite{DiLuzio:2015oha} 
worth to be mentioned.  
On a more technical side, the determination of Landau poles requires a two-loop accuracy 
in those cases where there is an accidental cancellations in the one-loop 
beta function (cf.~\fig{fig:2loopLP}). 
Moreover, in the classification of the effective operators in \Table{summary1} 
we pointed out the existence of a previously overlooked 
$d = 5$ operator responsible for a fast decay of the neutral component 
of the scalar multiplet $(1, 7, 0)$, effectively ruling out the minimal scalar DM candidate~\cite{Cirelli:2005uq}.

\section*{Acknowledgments}

L.D.L.~wishes to thank the organizers of the 27th Rencontres the Blois 
for providing a pleasant and stimulating atmosphere. 
This work was supported in part by the Slovenian Research Agency. 
The work of L.D.L.~is supported by the Marie Curie CIG program, project number PCIG13-GA-2013-618439. 
We thank the Galileo Galilei Institute for Theoretical Physics for the hospitality and the INFN for partial support 
during the completion of this work.

\end{document}